\documentclass[conference]{IEEEtran}

\newcommand{\ignore}[1]{}
\usepackage{fancyhdr}
\usepackage[normalem]{ulem}
\usepackage[hyphens]{url}
\usepackage{hyperref}
\usepackage{algorithm}
\usepackage[noend]{algpseudocode}
\usepackage{caption}
\usepackage{amsfonts}
\usepackage{subcaption}
\usepackage{booktabs, multicol, multirow}
\usepackage{lipsum}
\usepackage{dblfloatfix}
\usepackage{makecell}
\usepackage{graphicx}
\usepackage{tabularx}

\algrenewcommand\algorithmicindent{1.0em}%

\newcommand{\microsubmissionnumber}{396}

\fancypagestyle{firstpage}{
  \fancyhf{}
\setlength{\headheight}{50pt}

  \fancyhead[C]{\normalsize{MICRO 2017 Submission
      \textbf{\#\microsubmissionnumber} -- Confidential Draft -- Do NOT Distribute!!}} 
  \pagenumbering{arabic}
}  

\title{Snowflake: A Model Agnostic Accelerator \\for Deep Convolutional Neural Networks}

\author{\IEEEauthorblockN{Vinayak Gokhale\IEEEauthorrefmark{1},
Aliasger Zaidy\IEEEauthorrefmark{1}, 
Andre Xian Ming Chang\IEEEauthorrefmark{1},
Eugenio Culurciello\IEEEauthorrefmark{2}}
\IEEEauthorblockA{\IEEEauthorrefmark{1}School of Electrical and Computer Engineering}
\IEEEauthorblockA{\IEEEauthorrefmark{2}Weldon School of Biomedical Engineering}
Purdue University,
West Lafayette, IN 47907}

\begin{document}
\maketitle


\begin{abstract}

Deep convolutional neural networks (CNNs) are the deep learning model of choice for performing object detection, classification, semantic segmentation and natural language processing tasks. CNNs require billions of operations to process a frame. This computational complexity, combined with the inherent parallelism of the convolution operation make CNNs an excellent target for custom accelerators. However, when optimizing for different CNN hierarchies and data access patterns, it is difficult for custom accelerators to achieve close to 100\% computational efficiency.
In this work, we present Snowflake, a scalable and efficient accelerator that is agnostic to CNN workloads, and was designed to always perform at near-peak hardware utilization. Snowflake is able to achieve a computational efficiency of over 91\% on modern CNN models. Snowflake, implemented on a Xilinx Zynq XC7Z045 SoC is capable of achieving a peak throughput of 128\,G-ops/s and a measured throughput of 100 frames per second and 120\,G-ops/s on the AlexNet CNN model, 36 frames per second and 116\,G-ops/s on the GoogLeNet CNN model and 17 frames per second and 122\,G-ops/s on the ResNet-50 CNN model. To the best of our knowledge, Snowflake is the only implemented system capable of achieving over 91\% efficiency on modern CNNs and the only implemented system with GoogLeNet and ResNet as part of the benchmark suite.

\end{abstract}

\section{Introduction}

Deep convolutional neural networks (CNNs) have em-erged as the deep learning model of choice for object detection, classification \cite{alexnet_owt,googlenet,inception_v3,resnet}, semantic segmentation \cite{fcn_segmentation,resnet_segmentation}, natural language and speech processing tasks \cite{conneau_nlp}.
The success of deep learning can be attributed to two factors. The first is the increasingly large amount of computational power available in general purpose processors and the second is the availability of large datasets to train these networks. Early CNNs were designed for comparatively simpler tasks such as classification of digits \cite{lecun_lenet}. More challenging tasks such as being able to classify thousands of different categories and pixel-wise labeling required deeper CNNs. Over the past decade, the computational power necessary to train large networks in a reasonable amount of time was made available to researchers by means of general purpose graphics processing units (GPGPUs). This in turn resulted in a large number of CNN models with different network architectures, each one giving significantly better accuracy than its predecessor \cite{imagenet}. Recent CNN models are claimed to have better accuracy at performing classification in certain datasets than humans \cite{better_than_humans,inception_v4,resnet}.

The high computational complexity of modern CNNs coupled with the inherent parallelism of the convolution operator has resulted in a large number of custom accelerators that are more energy efficient than general purpose processors.
However, the highly varied data access patterns in CNNs makes it difficult for custom architectures to achieve high computational efficiency while processing a network. In this context, we define computational efficiency as the ratio of measured performance to peak performance. Alternatively, computational efficiency can also be expressed as a ratio of actual frames processed (actual fps) to the theoretical maximum number of frames that can be processed (peak fps). Most custom accelerators can either achieve high computational efficiency in some layers of a network but lower efficiency in other layers or have high efficiency for a particular CNN model but lower efficiencies for others.

In this paper, we present the SnowFlake CNN accelerator, which was designed with the primary goal of optimizing computational efficiency. We target computational efficiency because efficiency and bandwidth requirement are the two factors that affect throughput. For bandwidth-bound CNN layers, a variety of compression techniques have been presented which significantly reduce the size of these layers \cite{DeepCompression,SqueezeNet,han2016eie,han2015learning}. However, if the computations within a network do not map efficiently to the resources available in the system, computational efficiency will decrease. This is true for both uncompressed CNN models and models employing weight compression. Thus, maximizing computational efficiency is the motivation behind our design. We show that our design is able to achieve a computational efficiency of 95\% and can process networks in real time. Our design is scalable and can be used in a variety of systems from low powered embedded devices to large scale servers.

\section{Related Work}
\label{sec:related_work}
A variety of CNN accelerator designs have been proposed in recent years. Eyeriss is an ASIC implementation and uses a $12\times14$ grid of processing elements to accelerate CNN processing. The grid is fed by a 108\,KB scratchpad buffer. Eyeriss provides two performance figures for their design, one that includes DRAM load latency and one that does not. A case is made for the latter that DRAM latency is easy to optimize and hence can be ignored. Based on these two performance figures, Eyeriss achieves a computational efficiency on AlexNet of 68.8\% if DRAM load latency is included or 76.7\% if DRAM latency is ignored. On VGG (model D), Eyeriss achieves a computational efficiency of 31.8\% if DRAM latency is included or 36.5\% if DRAM latency is ignored. DianNao is an ASIC accelerator that focuses on optimizing memory accesses \cite{chen_diannao}. It is difficult to make a direct comparison with DianNao as they benchmark select layers of multiple networks while we benchmark entire networks. On the layers run on DianNao, they achieve 84.8\% efficiency when running a layer and reach a peak of 90.8\% towards the end of a layer's processing for a short period. The DianNao accelerator was improved and presented as ShiDianNao \cite{shidiannao}. A direct comparison with ShiDianNao is also difficult to make as they process smaller CNN models with layers that have a tens of kilobytes of weights and maps data and tens to hundreds of million operations per frame. By comparison, Snowflake is designed to accelerate networks with several megabytes of weights and maps data and billions of operations per frame.
The designs described in \cite{zhang_caffeine} and \cite{qiu_embedded_fpga} are FPGA-based accelerators capable of achieving 73.3\% and 80.3\%, respectively. By comparison, Snowflake achieves a maximum computational efficiency of 94\% on AlexNet.

\section{CNN Overview}
\label{sec:cnn_overview}

CNNs are feed-forward hierarchical models. The levels of hierarchy in a CNN are called the \textit{layers} of the network. A typical CNN model contains several layers. Each layer has three dimensional inputs called \textit{feature maps}. The outputs of one layer become the inputs to the next.

\begin{figure}[t]
  \centering
    \includegraphics[width=\columnwidth]{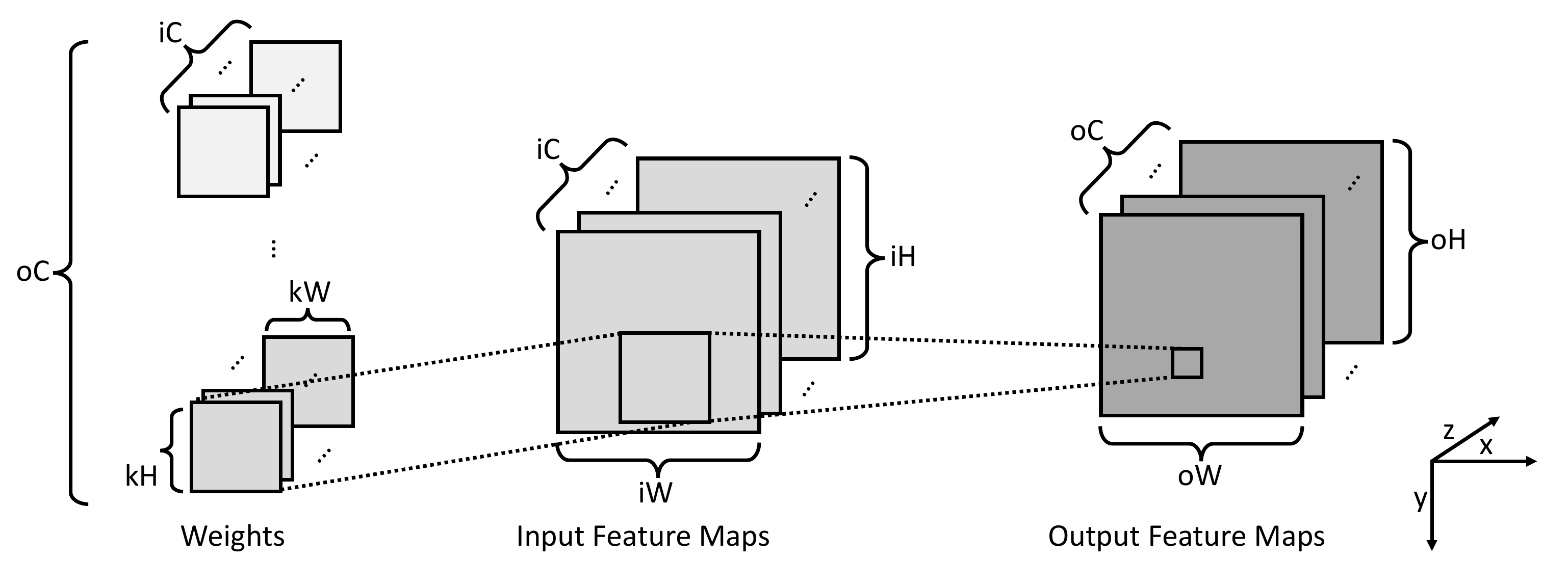}
    \caption{An overview of a CNN layer. The coordinate axis shown on the bottom left is used throughout this paper to depict dimensions.}
    \label{fig:cnn_overview}
\end{figure}

Figure \ref{fig:cnn_overview} shows the operation performed in one layer of a CNN. The layer shown in the figure has $iC$ input maps of width $iW$ and height $iH$ each. There are as many three dimensional convolutional kernels as output maps, which in figure \ref{fig:cnn_overview} is equal to $oC$. Each three dimensional kernel has dimensions $iC \times kH \times kW$. The elements of the convolutional kernels are called \textit{weights}. An output map is computed by first performing two dimensional convolution on each input map. The resulting $iC$ two dimensional outputs are added together by performing pixel-wise addition into a single two dimensional output feature map. This process is repeated by another three dimensional kernel $oC$ times to produce all the output feature maps. This operation is shown in equation \ref{eq:conv}.
\begin{equation}
\resizebox{.48 \textwidth}{!} 
{
$ofm(x,y,z_{o})= \sum^{iC}_{z_{i}=1}\sum^{kH}_{k_{y}}\sum^{kW}_{k_{x}}ifm(x+k_{x},y+k_{y},z_{i}) \times w(x,y,z_{i},z_{o})$
}
\label{eq:conv}
\end{equation}

where $ofm(x,y,z_{o})$ and $ifm(x+k_{x},y+k_{y},z_{i})$ are the output and input feature maps, respectively, and $w(x,y,z_{i},z_{o})$ are the four dimensional weights of the layer.

Some convolutional layers are followed by a downsampling operation. This serves to reduce the computational complexity in the network and provide translational invariance to the input. The most commonly used downsampling operation is called spatial max pooling. Max pooling performs the max operation over a two dimensional $P\times P$ window and produces a single output pixel. If the window has a stride of $s$ in both dimensions, it results in a downsampling of the feature maps by a factor of $s^{2}$. Some networks use average pooling instead of (or along with) max pooling.

Finally, regardless of whether pooling was applied or not, each convolutional layer is followed by a non-linearity. Early CNN models used the hyperbolic tangent function for the non-linearity. More recently, the rectified linear unit (ReLU) has become the non-linear function of choice, primarily because it facilitates faster training of CNN models \cite{alexnet}. ReLU sets negative inputs to zero and keeps non-negative inputs unchanged.

The outputs of the last convolutional layer are provided to a classifier. The first layer of the classifier converts these inputs to a vector and uses a matrix of weights to produce an output vector. A classifier can have multiple such layers. The classifier can be thought of as a convolutional layer with $1 \times 1$ convolutional kernels. The classifier is commonly also referred to as a fully connected (FC) layer because the input vector's elements are fully connected to all output vector elements by way of weights. FC layers are usually bandwidth dependent because they have comparatively lower computational complexity and have no data reuse in the weights, which make up the bulk of input data for the layer. In this paper, we primarily focus on the convolutional, max pooling and non-linear layers because they make up the majority of computations in a CNN. FC layers can be accelerated by using the same microarchitecture. However, a faster memory or compression techniques are often necessary to speed up the computation time required by these layers \cite{han2016eie}.

\subsection{Characteristics of CNN models}

CNNs are an active area of research and layer characteristics vary widely from one model to the next. From the point of view of compute, convolutional layers can be classified into three categories - conventional layers, inception modules and residual modules.

\paragraph{Conventional layers}

Conventional layers are of the type found in the Alex-Net \cite{alexnet_owt} and VGG \cite{vgg} models. These layers are similar to the layer shown in figure \ref{fig:cnn_overview} with $iC$ input maps, $oC$ output maps and $oC$ three dimensional kernels, each of which has dimensions $iC\times kH \times kW$.

\paragraph{Inception Modules}

Inception modules were introduced in the GoogLeNet model \cite{googlenet}. The major difference between Inception modules and conventional layers is that Inception modules use different kernel sizes on the same input maps volume. An inception module requires a three dimensional input of dimensions $iC\times iH\times iW$, just like conventional layers. However, it first applies three different groups of $iC\times1\times1$ kernels on this volume to produce three different branches of output map volumes. Two of these volumes have two differently sized kernels applied to them - the first has $3\times3$ kernels applied and the second uses $5\times5$. The depth of the kernels is equal to the number of maps in that branch. Finally, the original input volume with $iC$ input maps has max pooling applied to it. The output of the max pooling layer then undergoes a $1\times1$ convolution. This results in four separate output volumes which are then appended to produce the final output of the Inception module.

\paragraph{Residual modules}

Residual modules were introduced in \cite{resnet}. Residual modules have two forward paths. The first path is similar to a conventional layer. The input volume can have a large number of input maps - up to 2048 have been experimented with in \cite{resnet}. A layer called $1 \times 1$ reduce uses this large volume as input and produces fewer output maps - up to 512. These are then convolved with a larger kernel, typically $3 \times 3$. A $1 \times 1$ expand is performed on the output of this layer. The expand operation does the inverse of reduce by using a smaller number of maps as inputs and produces a larger number of outputs. At this stage the second path of the residual modules is used. This second path contains the volume that was used as input to the first path (and, by extension, the residual module). The number of output maps produced as the output of the first path is equal to its inputs. The second path is then added using element-wise addition to the output of the first path to produce the output of the residual module.

\section{Data Organization}
\label{sec:data_org}

Efficient data organization in memory is crucial for achieving peak performance. This section will describe the layout of maps and weights data in memory and help understand the motivation behind certain design decisions of the Snowflake microarchitecture. Snowflake organizes data in the form of \textit{traces}. A trace is defined as a contiguous region in memory that an instruction performs an operation on.

The primary motivation behind the concept of traces is to hide latencies of non-compute instructions behind compute cycles. By hiding the latencies of branches, true dependencies and memory loads and stores, Snow-flake is able to achieve close to peak performance. Due to the nature of a layer's inputs, there are three dimensions along which we can iterate first to produce an output value. Two of the three dimensions, the width and height of the kernel, are usually identical and only vary in whether data is accessed row-major or column-major. These dimensions are also usually the shorter of the three. The largest kernel size is found in AlexNet's layer 1 \cite{alexnet_owt} and is $11 \times 11$. By comparison, the third dimension, the depth of the input volume, can be as large as 2048 \cite{resnet}. In fact, apart from the first layer whose depth is 3, other layers have a depth of 32 or more \cite{alexnet_owt,googlenet,inception_v3,resnet,inception_v4}. Data organized in this \textit{depth-minor} format results in the longest trace. Based on the figure \ref{fig:cnn_overview}, if $iC = 256$ and $kW = kH = 3$, the trace length for this layer will $256 \times 3 = 768$. To produce one output element, $kH = 3$ such traces will be required. In the event that a single functional unit processes this trace, it will take 768 cycles to complete one trace. If a single instruction can keep this functional unit busy for 768 cycles, the instruction pipeline can proceed with other instructions necessary to prepare the second trace instruction while the functional unit is processing the first trace. The concept of traces can be extended to other compute instructions like max pool and non-compute instructions like loads and stores. This is the primary latency hiding technique used by Snowflake. In Snowflake, instructions that operate on traces are called \textit{vector instructions}. The longest and shortest trace lengths for some popular CNN models are shown in table \ref{tab:trace_len}.
\begin{table}[t]
  \centering
  \caption{The lengths of the longest and shortest traces when data is organized row- and column-major and depth-minor.}
  \resizebox{\columnwidth}{!}{
    \begin{tabular}{ccccc}
         & \multicolumn{2}{c}{Naive implementation} & \multicolumn{2}{c}{Depth-minor} \\
    Model & Longest trace & Shortest trace & Longest trace & Shortest trace \\
    \midrule
    AlexNet & 11    & 3     & 1152  & 33 \\
    VGG-D & 3     & 3     & 1536  & 9 \\
    GoogLeNet & 7     & 1     & 1024  & 21 \\
    ResNet-50 & 7     & 1     & 2048  & 21 \\
    \end{tabular}%
    }
  \label{tab:trace_len}%
\end{table}%

\section{Microarchitecture}

Due to the large number of multiply-accumulate (M-AC) operations involved in a CNN, Snowflake's design goal was primarily to accelerate these operations. Other operations include comparisons for performing maxpool-ing and activation functions. We describe the ReLU activation function in this paper. Functions like sigmoid and hyperbolic tangent can be added as piece-wise linear models as was done in \cite{nnx_by_moi,farabet_neuflow_somehow_still_novel}.

The Snowflake microarchitecture has two primary components - the control core which is responsible for generating instructions and the compute core which is responsible for processing a model's layers.

\subsection{Control Core}

The control core is similar to the five stage RISC pipeline with certain modifications made to fit Snow-flake's application specific nature. The control core provides vector instructions to the compute core. 

\paragraph{Instruction Fetch}

The instruction fetch stage contains the instruction cache and a program counter (PC) register. Snowflake's design makes it very good at hiding latencies of non-compute instructions. As a result, we did not require the use of techniques such as loop unrolling to improve throughput. This results in a very small code footprint. In our tests, the longest branch was always to an instruction within 512 instructions from the branch instruction. Snowflake's instructions are 32 bits wide. This required a minimum instruction cache size of 2\,KB. The instruction cache was double buffered so as to overlap the latency of fetching the next instruction stream, resulting in a 4\,KB instruction cache.

\paragraph{Instruction Decode}

The instruction decode stage is similar to the decode stage of the traditional RISC pipeline. This stage also performs a check for true data dependencies and stalls the fetch of further instructions until the dependent instruction commits.

\paragraph{Instruction Dispatch}

The instruction dispatch stage reads the source oper-ands from a register file. The dispatch stage also has hardware to keep track of the number of loads issued to the on-chip buffers of the compute core. This is necessary in order to prevent a vector instruction from reading data from these buffers while a load is pending, thus preventing read-after-write hazards through these buffers. 

\paragraph{ALU}

The ALU supports a subset of operations available in most other ALUs. It contains a multiplier, an adder and a comparator. It should be noted that the ALU is not directly involved in processing a CNN layer. Its purpose is to support bookkeeping operations necessary for issuing vector instructions to the compute core, which is where CNN processing occurs. For example, the adder in the ALU is utilized to increment the address of the compute core's buffers from where the next trace is to be read.

\paragraph{Register File}

The register file contains thirty two 32-bit registers and is dual ported. Data produced by the ALU is written to the register file if it has a write-back signal associated with it. Up to two source operands are read from the register file in the dispatch stage of the instruction pipeline.

\subsection{Compute Core}

\begin{figure*}[ht]
    \includegraphics[width=\textwidth]{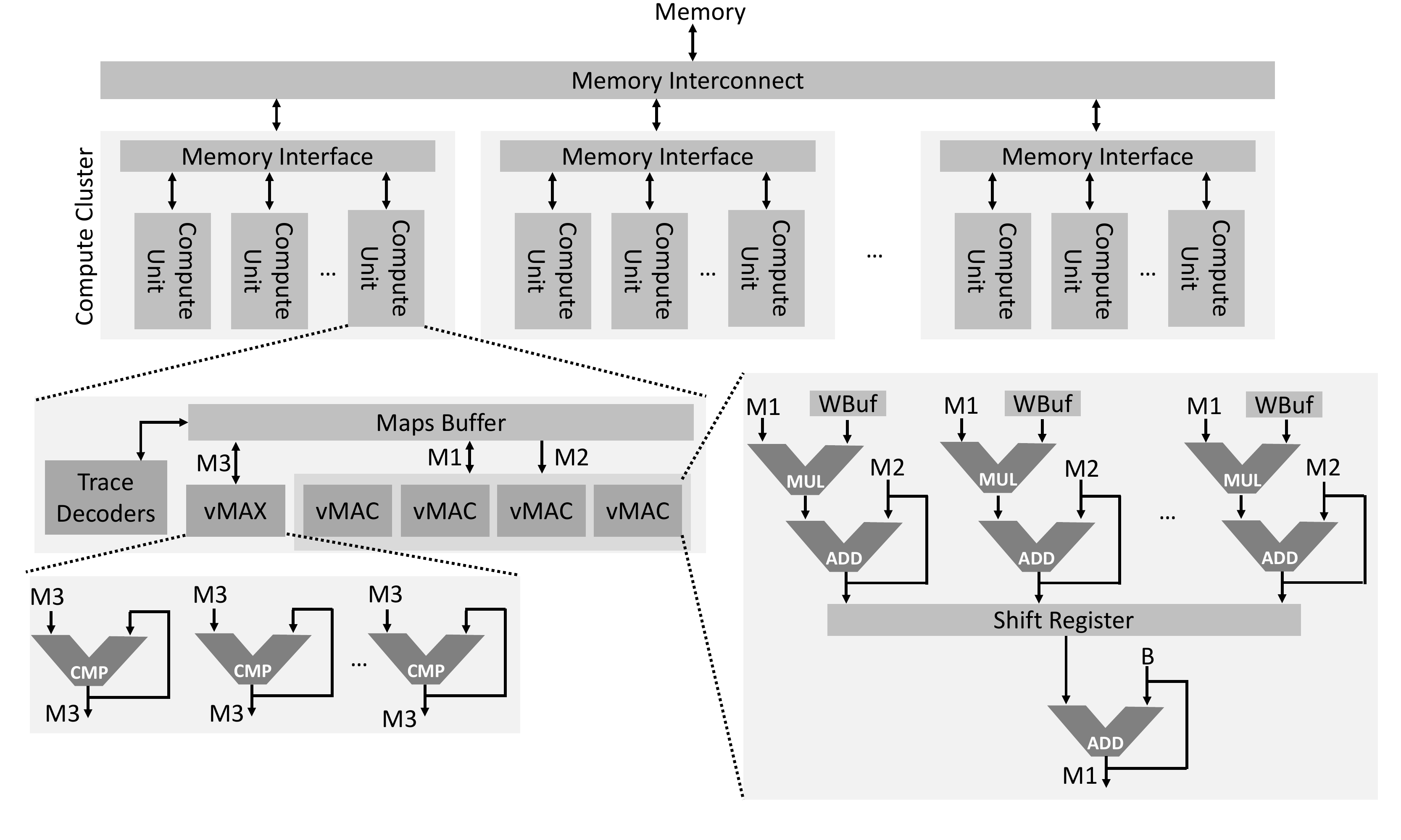}
    \caption{Snowflake has a hierarchical design and is composed of multiple compute clusters. Each cluster is composed of four compute units. Each compute unit contains four vMAC functional units and vMAX functional units, a maps buffer, weights buffers and trace decoders.}
    \label{fig:comp_cluster}
\end{figure*}

The compute core is responsible for processing the billions of operations required in the forward pass of a CNN model. The compute core is made up of multiple compute clusters and has a hierarchical structure, as shown in figure \ref{fig:comp_cluster}. Each compute cluster contains four compute units (CUs) and a memory interface. The compute units contain four vector multiply-accumuate (vMAC) units, each with 16 MAC units, one vector max pool (vMAX) unit, the trace decoders and a scratchpad memory called the maps buffer.

\paragraph{Trace Decoders} 
The trace decoders receive vector instructions from the control core. Vector instructions from the control core are pushed into a FIFO queue and are issued when their trace decoder is ready to receive an instruction. These instructions have the start address of the trace in the maps buffer, the length of the trace and the ID of the consumer. Based on the consumer ID, the instruction is forwarded to the appropriate trace decoder.
The decoders increment the start address and request a line from the maps buffer, once per cycle, until the length of the trace is reached. At this point, the next trace instruction is fetched and the process continues. The control core is responsible for having the next vector instruction available before the current one is complete. 

\paragraph{MAC Trace Decoder}
The MAC trace decoder fetches cache lines from the maps buffer and forwards them to the vMACs. It also provides addresses to index into the weights buffers inside each vMAC. This decoder is also responsible for signaling the vMACs to output the result accumulated in their internal accumulate registers. This signal is sent along with the last address of the final trace of the computation.

\paragraph{MAX Trace Decoder}
The MAX trace decoder fetches cache lines from the maps buffer and forwards them to the vMAX functional unit. Similar to the MAC trace decoder, it signals the vMAX unit to output the compared result after the final comparison.

\paragraph{Trace Move Decoder}
The trace move decoder is responsible for moving a trace from the maps buffer of one CU to the maps buffer of another CU (CU trace move) or storing a trace to memory (memory trace move). In case of a CU trace move, the destination CU has to be within the same cluster as the source CU. The logic implementing these two functions shares access to the maps buffer. If both types of trace move instructions are issued, the decoder will alternate between the two functions every cycle.

All three trace decoders continue processing their vector instructions, allowing the control core to proceed with scalar instructions that prepare the next set of vector instructions. Strictly speaking, the trace decoders enable Snowflake to execute \textit{and commit} vector instructions out of order with respect to scalar instructions. Correctness is maintained because of the guarantee that a scalar instruction never reads data produced by a vector instruction. This allows the control core to have a much simpler design than pipelines with speculative execution. Scalar (vector) instructions execute and commit in order with respect to other scalar (vector) instructions.

\subsubsection{Vector Multiply-Accumulate Units}

A compute unit's multipliers and adders are tied together into a multiply-accumulate (MAC) unit, as shown in figure \ref{fig:comp_cluster}. The multipliers work on 16-bit operands and the 32-bit result is connected to one of the inputs of the adder. The second operand of the adder is selected from either the adder's own output, or from a third operand. This third operand can be a partially accumulated result or an activation from a previous layer, as is used in residual modules \cite{resnet,inception_v4}. Snowflake uses 16-bit operands because prior work has shown that 16-bit fixed-point resolution has negligible impact on detection accuracy as compared to floating point numbers \cite{reduced_precision_gupta,chen_eyeriss_isca,chen_diannao}.

\begin{figure*}[t]
  \begin{subfigure}[t]{\columnwidth}
    \includegraphics[width=\textwidth]{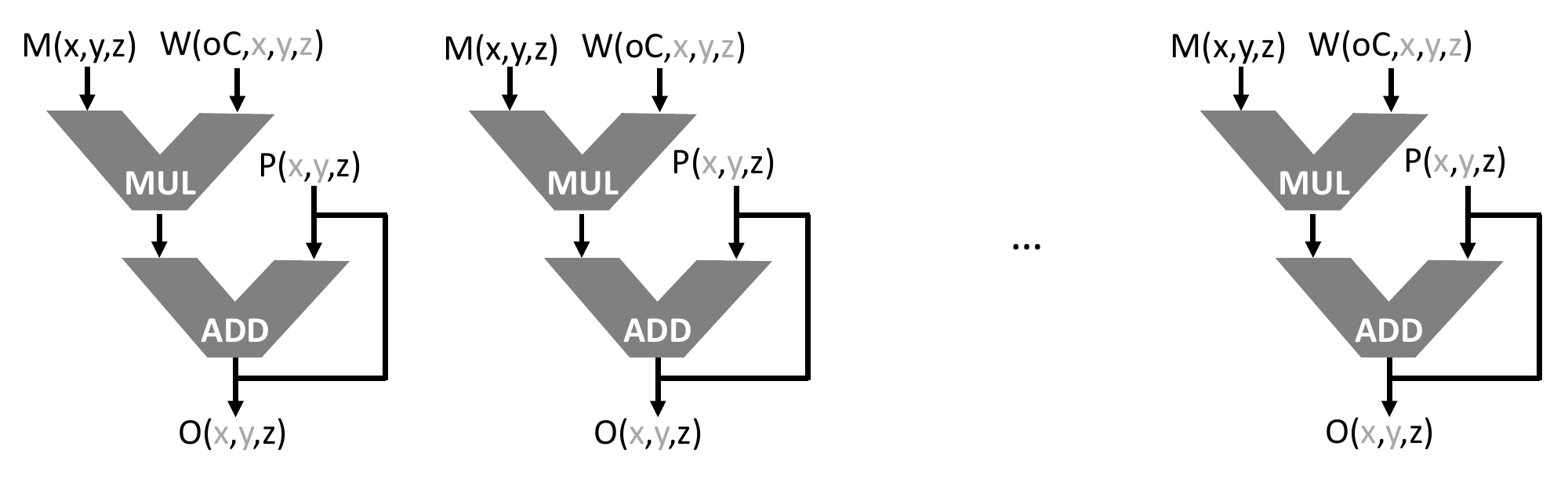}
    \caption{In the independent mode, each MAC in the vMAC operates independently to produce an output.}
    \label{fig:coop_mode}
  \end{subfigure}
  \hfill
  \begin{subfigure}[t]{\columnwidth}
    \includegraphics[width=\textwidth]{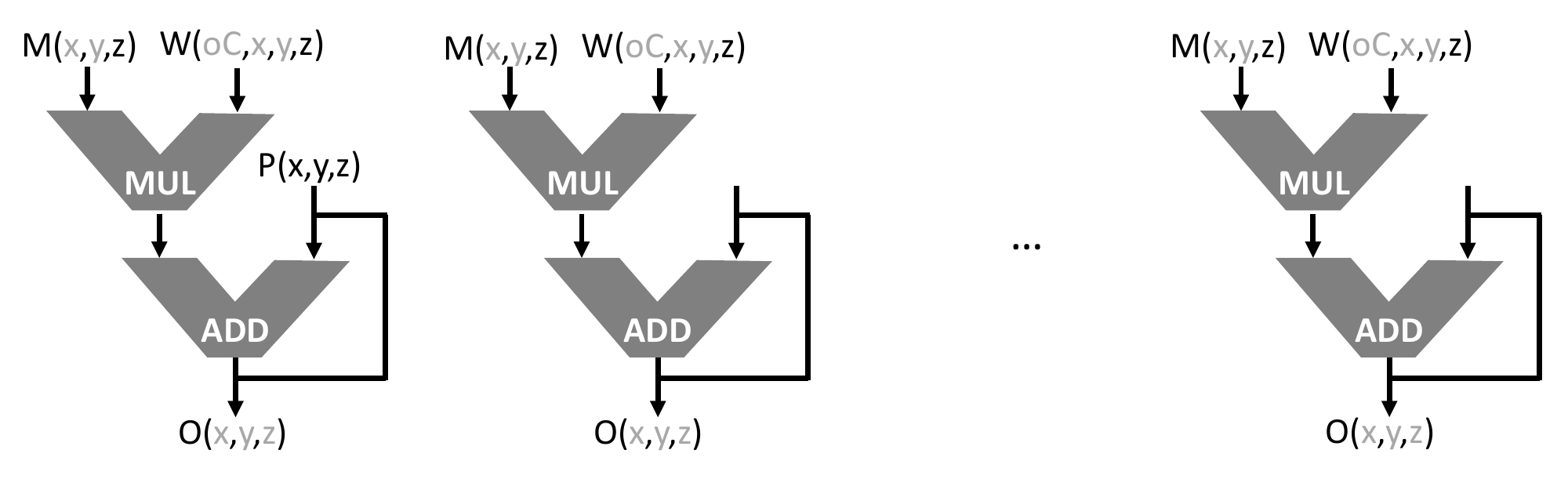}
    \caption{In the cooperative mode, the MACs within the vMAC work together to produce an output.}
    \label{fig:indep_mode}
  \end{subfigure}
  \caption{An illustration of the two modes of parallelism within the vMAC.}
  \label{fig:vmacs}
\end{figure*}

MAC units are organized in a group of 16 called a vector MAC unit (vMAC), as is shown in figure \ref{fig:vmacs}. In the figure, $M$ is a maps operand and $W$ is a weights operand. The $(x,y,z)$ coordinates are as shown in figure \ref{fig:cnn_overview} while $oC$ corresponds to the multiple three dimensional kernels that produce the output volume. The figure shows the two types of parallelism that can be exploited in Snowflake. The coordinates that vary across the MACs in a vMAC are shown in black. The coordinates that stay constant are shown in gray. Figure \ref{fig:indep_mode} shows \textit{inter}-output parallelism, where the output $O$ across the MACs has the same row and column location in the output volume but each MAC produces a different output map. This mode of computation is called the \textit{independent mode} (INDP) because every MAC works to produce an output independent of the others. In the INDP mode, each MAC gets the same input maps operand. The cache line containing the operand is fetched from the maps buffer based on the address provided by the MAC trace decoder. The operand is selected using the word address. Theoretically, one word from a cache line of 16 can be selected by means of a multiplexer in a single cycle. However, to reduce hardware complexity, a shift register is used. The shift register shifts the cache line every cycle until the requested word is at the front of the register. It is then broadcast to all MACs. If the layer is irregular, this can result in a drop in computational efficiency. For example, if the fifth word in a cache line is requested, there will be four cycles of latency before the MACs are supplied with the operand they need to continue processing the layer. However, due to the spatial locality inherent in the input maps volume, there is a high probability that the sixth word in the same line is the next one requested. This would be available the following cycle, and from then on until the next trace, there would be no latency. This type of irregularity is only encountered in layers with input maps that are not a multiple of 16.
Another inefficiency while using the INDP mode can result because of the fact that in INDP mode every MAC within the CU produces a different output map and there are 64 MACs in a CU. Due to this fact, the INDP mode requires that for achieving peak efficiency, there be a multiple of 64 output maps in the layer being processed. If there are fewer output maps, some MACs would need to be turned off resulting in a drop in efficiency.

Figure \ref{fig:coop_mode} shows data input when exploiting \textit{intra}-output parallelism. Every MAC within the vMAC receives an input from the maps volume that has the same row and column index but is from a different input map. The coordinates of the outputs are shown in gray because they are not independent from each other. All outputs produced by the MACs in this mode are partial results belonging to the same final output. Since the MACs work together to produce a single output, this mode is called the \textit{cooperative mode} (COOP) of computation. The partial results are accumulated together using a reduce operation that is performed by a separate adder. This adder, called the gather adder, is shown in figure \ref{fig:comp_cluster}. The partial results from the vMAC are latched into a shift register which feeds one partial result per cycle to one of the gather adder's inputs. The output from the gather adder is truncated to 16 bits and written back to the maps buffer. The address to write to is provided with the trace instruction. The COOP mode adds certain restrictions on what type of layers can be processed at peak efficiency. The gather adder requires 16 cycles to reduce the partial results produced by the MACs. If the MACs take fewer than 16 cycles to process a result, they would still need to wait until 16 cycles have passed before producing the next set of partial products. This requirement implies that the sum of the lengths of all traces required to produce an output using COOP mode should be 256. This restriction prevents layers that have 128 input maps and $1\times1$ convolution kernels from achieving peak efficiency using COOP mode. Such layers however can be run using the INDP mode at peak efficiency if there are at least 64 output maps. In either computation mode, the gather adder is also used to add the bias to the output.

The decision to have 16 MAC units per vMACs was made because of the restrictions imposed by the gather adder and the common characteristics of CNN layers. The gather adder requires as many cycles as MACs in a vMAC to add bias to (INDP) or reduce (COOP) the outputs of the MACs. Secondly, having the number of MACs equal to a power of two is necessary to optimally map the input maps to the MACs because the number of input maps is usually a power of two \cite{alexnet_owt,googlenet,inception_v3,resnet,inception_v4}. Increasing the number of MACs in a vMAC to 32 would result in the MACs in COOP mode processing the input traces twice as fast, while doubling the number of partial results the gather adder has to reduce. Inception and bottleneck modules frequently have $1\times1$ convolutions with 256 or 512 input maps which would require the INDP mode if the number of MACs within a vMAC was 32 or higher. With INDP mode, however, there is the restriction that the number of output maps has to be greater than the number of MACs within a CU, which is not the case for these Inception and bottleneck modules. This would result in decreased efficiency. 

On the other hand, decreasing the number of MACs per vMAC to 8 or lower holds no advantage based on the network topologies we have seen so far. It would double the latency available to the gather adder in each vMAC while also doubling the number of results produced per CU. However, it would also double the number of gather adders resulting in additional hardware that is unnecessary given the current trends in CNN model architecture. Theoretically, a single gather adder could be shared by  two vMACs, thereby keeping the hardware cost the same as a 16 MAC vMAC, but this would require extra routing and control logic while providing little benefit.

\subsubsection{Vector Maxpool Units}

Four 16-bit comparators are grouped together into a vector Maxpool (vMAX) unit. The vMAX unit processes 16 words at a time. Each comparator is provided four words and requires 4 cycles to process the words. Hence, a 3x3 maxpool window requires 3*3*4 = 36 cycles to complete and produces 16 words.

Since the vMACs can write data back to the maps buffer, processing of a maxpool layer can proceed without requiring access to main memory. The control core can issue vector instructions to the vMACs and vMAX functional units simultaneously, thereby allowing Snow-flake to hide the maxpool processing latency behind the MAC operations. The results from the vMAX units are stored back to the maps buffer.

\subsubsection{Scratchpad Memory}

The scratchpad memory within a CU is split into two parts - the maps buffer and weights buffers. There are as many weights buffers as MACs within the design. Each MAC has a weights buffer connected to one of its inputs, as shown in figure \ref{fig:comp_cluster}. Depending on the mode of computation, one or all of the weights buffers within a vMAC also provide the bias operand. In COOP mode, only one bias operand is necessary as all MACs produce partial results of the same output. In INDP mode, all weights buffers provide the bias operand. The bias value remains constant throughout the entire output map. When a new output map is being computed, the bias is loaded once at the beginning into a register connected to the second operand of the MAC unit's adder. The MAC trace decoder provides the weights buffer with the address at which the weights operand is located.

\begin{figure}[!b]
  \includegraphics[width=1\columnwidth]{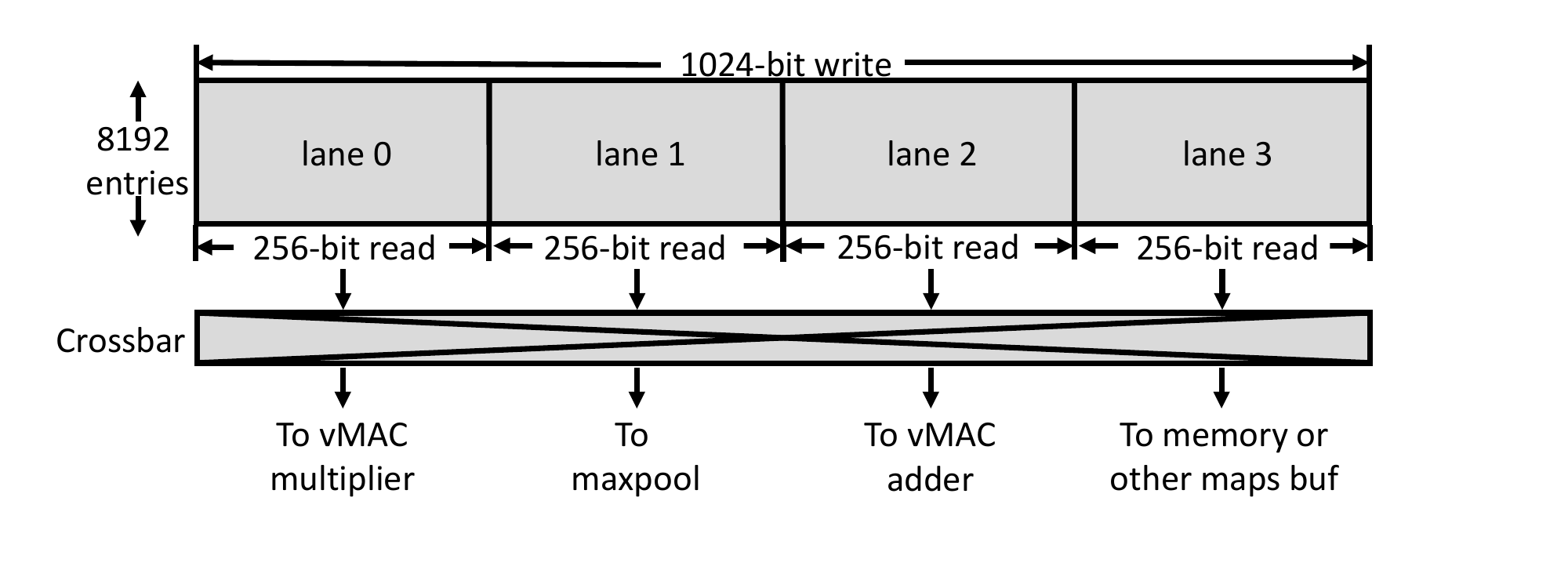}
  \caption{The maps buffer has one 1024-bit write port and allows up to four different 256-bit reads to proceed per cycle.}
  \label{fig:m_buf}
\end{figure}

The organization of the maps buffer is shown in figure \ref{fig:m_buf}. The maps buffer has a 1024-bit write port and four banks, each with 256-bit read ports called \textit{lanes}. vMACs consume an entire 256-bit cache line in one cycle if they are operating in COOP mode. They require up to 16 cycles to consume a line if operating in INDP mode.

Data is written to the maps buffer from memory, the vMACs, the maxpool comparators or from another CU's maps buffer. The maximum bit width is 1024 but there are enable signals at a 64-bit granularity. This is to allow data from the vMACs, which for COOP mode is 64 bits, to be written back.

All three trace decoders can access the maps buffer every cycle. The fourth port is used to access a third operand and can be used to access the bypassed connections in residual modules or for previously computed partial results. The MAC trace decoder gets priority to read from the buffer. Its read request to any lane is always granted. The read for the other decoders is granted only if they do not access the same lane as the MAC decoder. The trace move decoder's access is shared by its memory-move and CU-move functions. If both memory- and CU-move instructions are issued, they alternate accesses to the buffer. The read request from the decoders is given to the source lane. The lane is identified by using the lower two bits of the read address. Cache lines read from the lanes are directed towards their requestors.

\subsection{Instruction Set}
\label{sec:isa}

Snowflake instructions are 32 bits wide and are designed around the concept of traces introduced in section \ref{sec:data_org}. Instructions have a 4-bit operand code and most instructions use a fifth bit called the \textit{mode bit}. The mode bit is used to distinguish behavior within a particular of instruction. Snowflake's instructions are divided into four types - data move, compute, branch and memory access instructions. Some types have both scalar and vector instructions. Both scalar and vector instructions use the general purpose registers as the source. Scalar instructions also use the registers as the destination. The destination for all vector instructions is the maps buffer. 

Apart from the general purpose registers, there are a set of registers, one per CU that control the write-back address for the MAC and MAX instructions. Since the MAC and MAX values are produced in a strided pattern, the write-back addresses can be computed by a base address and offset pair. Every MAC trace instruction that results in a write-back increments the base address by the offset. The address value is sent along with the vector instruction to the trace decoders. This eliminates the need for a dedicated store instruction. Data is moved into these registers by a data move instruction.

\subsubsection{Data Movement Instructions}

The scalar data move instructions (MOV) move data from the immediate field into a register (mode 0) or from one register to another (mode 1). In mode 1, MOV instructions also include a 5-bit shift field that allows the value in the source register to be left-shifted by up to 32.

Vector data move instructions are of two types. The first is a trace move (TMOV) which moves a trace of up to 4096 words from the maps buffer of one CU to the maps buffer of another CU. The second is a vector move (VMOV) that moves one 256-bit cache line from the maps buffer to the registers that feed the multiply-accumulate units.

\subsubsection{Compute Instructions}

Scalar compute instructions are the addition and multiplication instruction. For both instructions, mode 0 uses one source operand and an immediate value while mode 1 uses two source operands.

The vector compute instructions are the MAC and MAX instructions. One of the source registers of the MAC instruction contains the start address of the maps trace while the second source register contains the start address of the weights trace. The immediate value provides the length of the trace. Mode 0 for the MAC instruction signals the vMACs to operate in INDP mode while mode 1 is for COOP mode. For the MAX instruction, only one source register is used which provides the start address for the maps trace. MAX instructions do not require access to the weights buffers and only have one computation mode.

\subsubsection{Branch Instructions}

There are three branch instructions - branch if greater than, branch if less than or equal to and branch if equal to. The two source registers provide the two values to be compared. The immediate value contains the offset from the current PC at which to branch to. A comparator in the ALU of the control core resolves the branch and the program counter is updated if the branch is taken.

A branch instruction is followed by four branch delay slots. The decision to include branch delay slots was taken to avoid more complicated techniques like branch prediction. Branch delay slots also avoid the need to include logic that flushes the stages of the control core in the event that a branch is taken. Finally, most commonly encountered branches in a CNN layer can make use of at least 3 of the four delay slots.

\subsubsection{Memory Access Instructions}

Memory access instructions are loads and stores, both of which are vector instructions. In the load instruction, one source register contains the start address of the trace in memory. The second source register contains two fields. The lower 23 bits contain the start address of the weights or maps buffer where the trace that is fetched is to be stored. The upper 9 bits are used to identify the destination buffer. 4 of the bits specify the CU while the other 5 specify the buffer ID within a CU.

The store instruction writes a trace from the maps buffer to memory. In the store instruction, the first source register contains the memory address where the first word in the trace it to be stored. The second source register contains the maps buffer address of the trace and bits for selecting the target CU within the compute cluster. The immediate value in the instruction specifies the trace length.

\section{Results}
\subsection{System Specifications}
We implemented Snowflake with 4 CUs (1 compute cluster) on the Xilinx Zynq XC7Z045 device. The Zynq is an SoC that contains an FPGA and two ARM cortex-A9 processors on the same die. The platform used was the ZC706 development board. This board has the Zynq device, 1\,GB of DDR3 memory that is shared by the ARM cores and the FPGA. The characteristics of the system used in our experiments is shown in table \ref{tab:sf_specs}. The ARM cores are not involved in the processing of the layers. They are used to write the instruction stream to the shared DDR3 and initialize Snowflake with the PC pointing to the first instruction in this instruction stream. From there, Snowflake's control core takes over and is responsible for processing the model.

\begin{table}[h]
  \centering
  \caption{System specifications of the implemented Snowflake system.}
    \begin{tabular}{ll}
    \textbf{Platform} & ZC706 \\
    \midrule
    \textbf{Device} & Xilinx Zynq XC7Z045 \\
    \midrule
    \textbf{Processor} & 2x ARM Cortex A9 \\
    \midrule
    \textbf{Memory} & 1 GB DDR3 \\
    \midrule
    \textbf{Memory B/W} & 4.2 GB/s \\
    \midrule
    \textbf{MAC Units} & 256 \\
    \midrule
    \textbf{Accelerator Frequency} & 250 MHz \\
    \midrule
    \textbf{Peak Performance} & 128 G-ops/s \\
    \midrule
    \textbf{Power Consumption} & 9.5 W \\
    \bottomrule
    \end{tabular}%
  \label{tab:sf_specs}%
\end{table}

The 4 CU Snowflake system contains a 128\,kB maps buffer and four vMACs, each with a 16\,kB weights buffer. This results in a total on-chip memory of 768\,kB. 

\subsection{Experimental Results}

We benchmarked Snowflake's performance on a benchmark suite consisting of AlexNet \cite{alexnet_owt}, GoogLeNet \cite{googlenet} and ResNet-50 \cite{resnet}. We chose these models because each contains a different type of layer hierarchy. AlexNet is made up entirely of conventional layers, GoogLeNet contains Inception modules and ResNet-50 has residual modules. We did not feel the need to include VGG as a benchmark because its layer are similar to AlexNet's layers 3-5 and the network itself has a high computational complexity despite having a lower accuracy than both GoogLeNet and ResNet-50 \cite{googlenet,resnet}. 

\subsubsection{AlexNet}

In table \ref{tab:alexnet_layer_perf} we show layer-wise performance and efficiency for AlexNet. INDP mode is used for layer 1 and COOP mode is used for the other layers. The inefficiency in layer 1 comes from the irregularity of the $3 \times 11 \times 11$ kernels. These kernels result in a trace length of 33 which is not a multiple of the cache line size. This results in  unaligned accesses which incur a delay when shifting the line to fetch the requested word. This delay does not exist in COOP mode which results in an efficiency close to 100\% for layers 2-5. 
\begin{table}[htbp]
  \centering
  \caption{Layer-wise breakdown of performance on AlexNet.}
  \resizebox{\columnwidth}{!}{
    \begin{tabular}{cccccc}
   \begin{tabular}{@{}c@{}}\textbf{Layer} \\ \textbf{\#}\end{tabular}
   & \begin{tabular}{@{}c@{}}\textbf{Ops} \\ \textbf{(M-ops)} \end{tabular}
   & \begin{tabular}{@{}c@{}}\textbf{Theor.} \\ \textbf{Time (ms)}\end{tabular} 
   & \begin{tabular}{@{}c@{}}\textbf{Actual} \\ \textbf{Time (ms)}\end{tabular}
   & \begin{tabular}{@{}c@{}}\textbf{Perf.} \\ \textbf{(G-ops/s)}\end{tabular}
   &  \begin{tabular}{@{}c@{}}\textbf{Eff.} \\ \textbf{\%}\end{tabular} \\
   \midrule
    1     & 139 & 1.09  & 1.56 & 87.4  & 69.87 \\
    2     & 409 & 3.19  & 3.22 & 127.0 & 99.07 \\
    3     & 202 & 1.58  & 1.59 & 127.0 & 99.37 \\
    4     & 269 & 2.10  & 2.16 & 126.3 & 97.22 \\
    5     & 179 & 1.40  & 1.42 & 125.2 & 98.59 \\
    \midrule
    \textbf{Total} & \textbf{1198.00} & \textbf{9.36} & \textbf{9.95} & \textbf{120.3} & \textbf{94.07} \\
    \bottomrule
    \end{tabular}
    }
  \label{tab:alexnet_layer_perf}
\end{table}

\begin{figure}[t]
  \includegraphics[width=1\columnwidth]{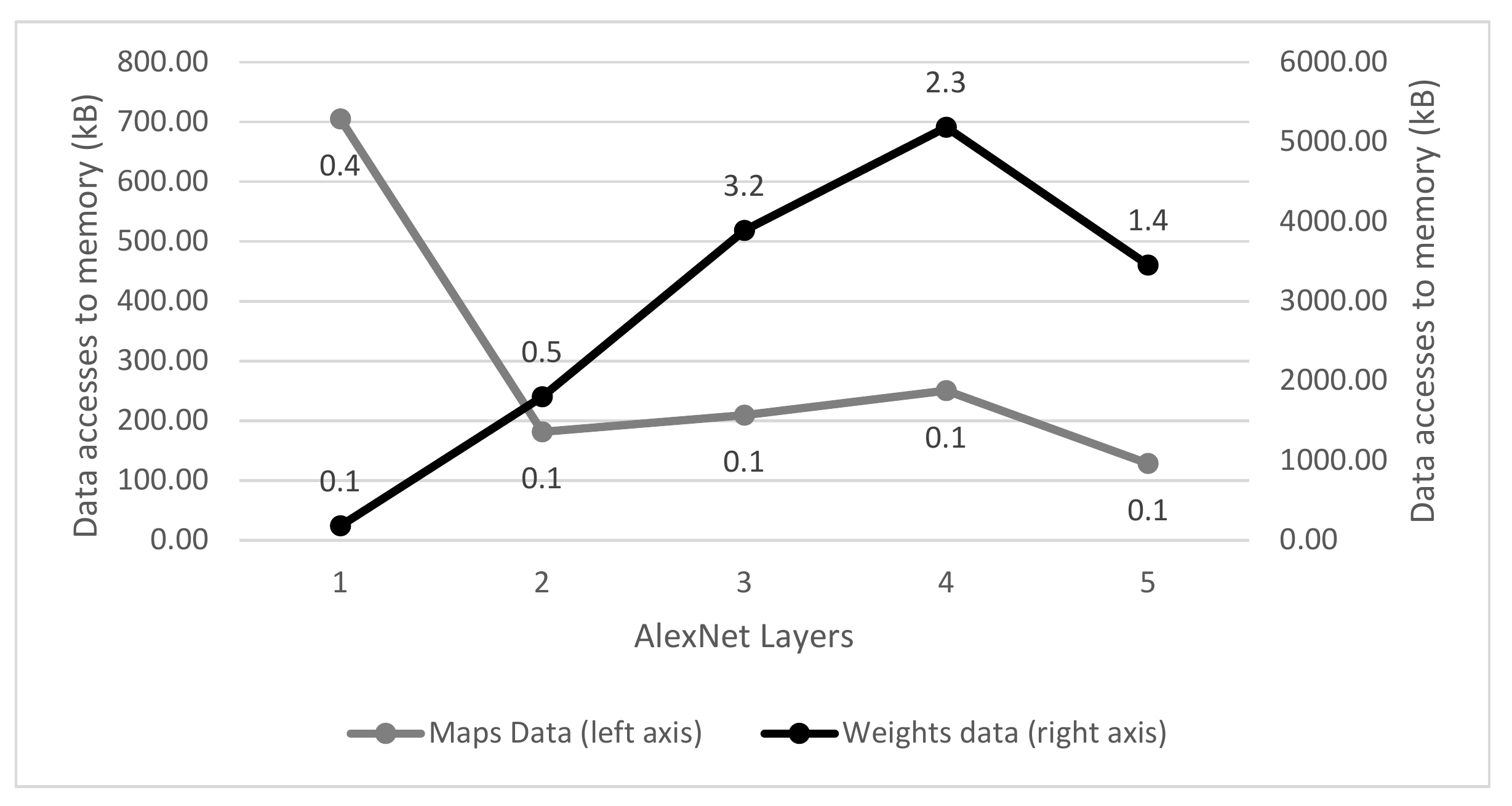}
  \caption{this figure shows the size of weights and maps data moved to and from memory. The left axis is for the maps and the right axis is for the weights. The numbers written above each point are the bandwidth in GB/s.}
  \label{fig:alexnet_bw}
\end{figure}

The layer-wise bandwidth requirements for AlexNet are shown in figure \ref{fig:alexnet_bw}. AlexNet requires an average bandwidth of 1.53\,GB/s. Layer 1 has the lowest bandwidth requirement at 0.27\,GB/s. This is because all weights fit in the weights buffers and all maps fit in the maps buffers. This does not require tiling either the weights or the maps volumes and is the best-case bandwidth possible. Layer 4 has the highest bandwidth requirement. In layers 2-5, the input maps volume is split into three tiles. The weights are cycled through the accelerator thrice. This accounts for the higher bandwidth in layers 3-5. For layer 2, the size of the weights volume is about half the size for the next layer but the time required to process this layer is double. As a result, the bandwidth is a quarter of the bandwidth required for the next three layers.

We also ran the five convolutional layers in AlexNet end-to-end. Throughput for the end-to-end run was similar to the sum of the layer-wise runs. This is made possible by efficiently double buffering the input maps volume and the instructions which removes any configuration latency between the layers. The FC layers were not included in the benchmarks. However, FC layers can be viewed as $1 \times 1$ convolutional layers and usually do not have any irregularity. As a result, given enough bandwidth, Snowflake can accelerator these layers at 100\% efficiency. Bandwidth required for the first FC layer in AlexNet given Snowflake's throughput of 128\,G-ops/s is 234\,GB/s. However, compression techniques as the ones described in \cite{han2016eie} can significantly reduce the bandwidth required. Compression can be included into Snowflake's design by adding modules that decompress data before writing them to the maps buffer.

\subsubsection{GoogLeNet}

The first two layers of GoogLeNet are conventional layers and the rest are Inception modules. In table \ref{tab:perf_modules_googlenet}, we present the results of layer- and module-wise runs on GoogLeNet. It should be noted that layer 2 is comprised of two parts, the first is a $1\times1$ convolutional layer whose input and output channels are 64. The second is a $3 \times 3$ convolutional layer with 192 output maps. 
On GoogLeNet, average computational efficiency was 91.5\%. Layer 1 had the lowest efficiency at 73.6\% and layer 2 had the highest efficiency of 97.3\%. Similar to AlexNet's layer 1, GoogLeNet's inefficiency for layer 1 comes from irregularity in the $3\times7\times7$ kernels. On the other hand, layer 2 is highly regular and requires 703 M-ops which is 22\% of the entire computational complexity of the model.
For GoogLeNet's inception modules, the lowest efficiency was 86.7\% for module 3a and the highest efficiency was 95.3\% for 3b and 5b. Inception 3a has 192 input maps and four branches. Recall that inception modules first reduce the input maps volume to multiple smaller volumes using $1\times1$ convolutional kernels to create multiple branches. The modules then use larger kernels on these branches. For inception 3a, the $1\times1$ reduce happens on 192 input maps. For such a layer, the trace length is 192 which is less than the 256 necessary to operate in COOP mode. 192 input maps are not irregular and always result in aligned accesses. However, one of the branches produces 16 output maps while another produces 96. For INDP mode to operate at peak efficiency, a multiple of 64 output maps are necessary. This results in the former branch operating at 25\% efficiency and the latter branch operating at 75\% efficiency. This accounts for the drop in efficiency for inception 3a. GoogLeNet's average efficiency of 91.5\% translates to a throughput of 113.2\,G-ops/s, a processing latency of 27.75\,ms and a frame rate equal to 36 frames per second.

\begin{table}[htbp]
  \centering
  \caption{Layer- and module-wise performance on the GoogLeNet model.}
  \resizebox{\columnwidth}{!}{
    \begin{tabular}{cccccc}
    \begin{tabular}{@{}c@{}}\textbf{Layer} \\ \textbf{\#} \end{tabular}
    & \begin{tabular}{@{}c@{}}\textbf{Ops} \\ \textbf{(M-ops)} \end{tabular}
    & \begin{tabular}{@{}c@{}}\textbf{Theor.} \\ \textbf{Time (ms)} \end{tabular}
    & \begin{tabular}{@{}c@{}}\textbf{Actual} \\ \textbf{Time (ms)} \end{tabular}
    & \begin{tabular}{@{}c@{}}\textbf{Perf.} \\ \textbf{(G-ops/s)} \end{tabular}
    & \begin{tabular}{@{}c@{}}\textbf{Eff.} \\ \textbf{\%} \end{tabular} \\
    \midrule
    Layer 1 & 236   & 1.84  & 2.50  & 94.4  & 73.7\% \\
    \midrule
    Layer 2 & 756   & 5.49  & 5.64  & 134.0 & 97.3\% \\
    \midrule
    Inception 3a & 256   & 2.25  & 2.59  & 98.9  & 86.9\% \\
    Inception 3b & 609   & 4.98  & 5.22  & 116.6 & 95.4\% \\
    \midrule
    Inception 4a & 147   & 1.28  & 1.45  & 101.5 & 88.3\% \\
    Inception 4b & 176   & 1.49  & 1.69  & 104.0 & 88.2\% \\
    inception 4c & 214   & 1.66  & 1.87  & 114.4 & 88.8\% \\
    inception 4d & 237   & 1.92  & 2.03  & 116.8 & 94.6\% \\
    Ineption 4e & 340   & 2.68  & 2.84  & 119.7 & 94.4\% \\
    \midrule
    Inception 5a & 112   & 0.78  & 0.83  & 134.9 & 94.0\% \\
    Inception 5b & 141   & 1.04  & 1.09  & 129.7 & 95.4\% \\
    \textbf{Total} & 3224  & 25.41 & 27.75 & 116.2 & 91.6\% \\
    \bottomrule
    \end{tabular}%
    }
  \label{tab:perf_modules_googlenet}%
\end{table}%

GoogLeNet has an average pool layer that follows Inception module 5b. This layer converts the $1024 \times 7 \times 7$ output volume of Inception 5b into a $1024 \times 1 \times 1$ vector. This vector is then provided as an input to a 1024-to-1000 FC layer. Average pooling can be viewed as a convolution with a kernel whose weights are all equal to a value that is the inverse of the average pooled area. For GoogLeNet's average pool layer, this value is $1 \div 49 = 0.02$. The average pool layer has low computational complexity - it requires only 98,000 operations to produce the output and a theoretical run time of 0.7\,microseconds. As a result, it has a very high bandwidth requirement which reduces its performance on Snowflake. Average pool has a theoretical efficiency of 100\% but is reduced to 23.3\% in our benchmark run due to this high bandwidth requirement. However, due to its low computational complexity, its low efficiency has a negligible effect on the efficiency and performance of the entire network.

\subsubsection{ResNet-50}

ResNet's 50 layer version has the highest computation cost among the three benchmarks. The first layer in this model is identical to GoogLeNet's first layer. The remaining layers are the bottleneck modules. The results of module-wise runs on ResNet-50 are shown in table \ref{tab:perf_resnet}. Each conv\_x module in the table has a bottleneck module replicated multiple times. 
\begin{table}[htbp]
  \centering
  \caption{Layer- and bottleneck module-wise performance on ResNet-50.}
  \resizebox{\columnwidth}{!}{
    \begin{tabular}{cccccc}
    \begin{tabular}{@{}c@{}}\textbf{Layer} \\ \textbf{\#} \end{tabular}
    & \begin{tabular}{@{}c@{}}\textbf{Ops} \\ \textbf{(M-ops)} \end{tabular}
    & \begin{tabular}{@{}c@{}}\textbf{Theor.} \\ \textbf{Time (ms)} \end{tabular}
    & \begin{tabular}{@{}c@{}}\textbf{Actual} \\ \textbf{Time (ms)} \end{tabular}
    & \begin{tabular}{@{}c@{}}\textbf{Perf.} \\ \textbf{(G-ops/s)} \end{tabular}
    & \begin{tabular}{@{}c@{}}\textbf{Eff.} \\ \textbf{\%} \end{tabular} \\
    \midrule
    conv\_1 & 232  & 1.81  & 2.76  & 84.1  & 65.7\% \\
    conv\_2 & 1165 & 9.10  & 9.37  & 124.4 & 97.2\% \\
    conv\_3 & 1857 & 14.51 & 14.93 & 124.4 & 97.2\% \\
    conv\_4 & 2388 & 18.66 & 20.55 & 116.2 & 97.0\% \\
    conv\_5 & 1235 & 9.65  & 10.63 & 116.2 & 97.0\% \\
    \midrule
    \textbf{Total} & 6879 & 53.72 & 56.25 & 122.3 & 95.5\% \\
    \bottomrule
    \end{tabular}
    }
  \label{tab:perf_resnet}
\end{table}

\begin{table*}[t]
  \centering
  \caption{A comparison of throughput and efficiency across recent works in literature.}
  \resizebox{\textwidth}{!}{
    \begin{tabular}{lccccccccc}
          & \multicolumn{2}{c}{\textbf{Eyeriss\cite{chen_eyeriss_jssc}}} & \textbf{Zhang\cite{zhang_CNN_optimized_FPGA_arch}} & \textbf{Caffeine\cite{zhang_caffeine}} & \textbf{Qiu\cite{qiu_embedded_fpga}} & \textbf{HWCE\cite{meloni}} & \multicolumn{3}{c}{\textbf{Snowflake}} \\
          \midrule
          & AlexNet & VGG   & AlexNet & VGG   & VGG   & AlexNet & AlexNet & GoogLeNet & ResNet-50 \\
    \textbf{Platform} & 65nm CMOS & 65nm CMOS & VX485T & KU060 & Zynq 7045 & Zynq 7045 & Zynq 7045 & Zynq 7045 & Zynq 7045 \\
    \textbf{Clock (MHz)} & 200   & 200   & 100   & 200   & 150   & 100   & 250   & 250   & 250 \\
    \textbf{Precision} & 16-bit fixed & 16-bit fixed & 32-bit float & 16-bit fixed & 16-bit fixed & 16-bit fixed & 16-bit fixed & 16-bit fixed & 16-bit fixed \\
    \textbf{MAC Units} & 168   & 168   & 448   & 1058  & 780   & 800   & 256   & 256   & 256 \\
    \textbf{Actual Perf. (G-ops/s)} & 46.1  & 24.5  & 61.6  & 310.0 & 187.8 & 140.8 & 120.3 & 116.2 & 122.3 \\
    \textbf{Peak Perf. (G-ops/s)} & 67.2  & 67.2  & 89.6  & 423.2 & 234   & 160   & 128   & 128   & 128 \\
    \textbf{Frame Rate (fps)} & 38.4  & 0.8   & 51.3  & 258.3 & 6.3   & 117.3 & 100.3 & 36.3  & 17.7 \\
    \textbf{Power (W)} & 0.28  & 0.24  & 18.61 & 25    & 9.63  & -     & 9.48  & 9.53  & 9.61 \\
    \textbf{Energy Eff. (G-ops/J)} & 164.6 & 102.1 & 3.3   & 12.4  & 19.5  & -     & 12.7  & 12.2  & 12.7 \\
    \textbf{Computational Eff.} & 69\%  & 36\%  & 69\%  & 73\%  & 80\%  & 88\%  & \textbf{94\%} & \textbf{91\%} & \textbf{95\%} \\
    \end{tabular}
    }
  \label{tab:perf_comparison}
\end{table*}

Each bottleneck module within a conv\_x module is identical. As a result, these were run only once. Additionally, conv\_2 and conv\_3 similar in terms of the hierarchy of the bottleneck modules. Due to this, we benchmarked the more computationally expensive conv\_3 and the performance for conv\_2 was extrapolated from this run. Similarly, conv\_4 and conv\_5 are hierarchically similar and conv\_5 was benchmarked while the performance for conv\_4 is an extrapolation. As is the case for the input layer of the first two benchmarks, ResNet's first layer is also irregular and has a lower efficiency. However, the bottleneck modules are all highly regular and have a large number of input maps, especially in conv\_3 to conv\_5. This results in high throughput for these modules. As a result, the efficiency of the entire network is 95.5\%. Snowflake is able to process ResNet-50 at a throughput of 17.7 frames per second.

\subsection{Performance Comparison}

We compare our performance to other recent works in literature in table \ref{tab:perf_comparison}. 
Eyeriss is an ASIC implementation \cite{chen_eyeriss_jssc} and has a grid of $14\times12$ MAC units. Eyeriss has a computational efficiency of 69\% and 36\% for AlexNet and VGG, respectively. While Eyeriss does run-length encoding of the input and output maps, they decompress the input maps volume before processing them in the grid. Computational efficiency numbers are not directly available in \cite{chen_eyeriss_jssc}. However, the paper does provide the means to compute both peak and measured performance. Peak performance is computed as $2\times numPEs\times clock$. The factor of 2 comes from counting one multiply-accumulate as two operations. Measured performance is computed by dividing operations for the model by the latency. Eyeriss provides two metrics for latency. Total latency is defined as the measured latency to process a layer. Processing latency is defined as total latency less time taken to fetch maps from DRAM and write outputs back to DRAM. As such, it is a simulated metric. However, they claim to be able to optimize this by better controlling DRAM traffic at negligible cost. In the interest of fairness, we have considered the simulated processing latency metric for performance numbers.

Zhang et.al. \cite{zhang_CNN_optimized_FPGA_arch} use a 32-bit floating point accelerator. While this paper lists the number of MAC units as 2280, we do not directly use the same method as above because a floating point MAC unit requires 5 fixed point MAC units, as listed in the paper. Thus, we divide the above number by 5 to get 448 as the number of floating point MAC units available in the design. We then use the above equation along with a frequency of 100\,MHz to get 89.6\,G-ops/s as the peak performance. The paper states measured performance as 61.6 G-ops/s which results in an efficiency of 69\%.

Caffeine is an accelerator implemented in the newer Xilinx Ultrascale FPGA architecture using the KU060 device \cite{zhang_caffeine}. The accelerator contains 1058 MAC units clocked at 200\,MHz which results in a peak performance of 423.2\,G-ops/s. However, Caffeine provides peak performance of 365\,G-ops/s. For consistency, we have used the former value of peak performance. Using the provided value of 310\,G-ops/s as measured performance for convolutional layers, we compute a computational efficiency of 72\%. If the value for peak performance provided in the paper is considered, it results in an efficiency of 84\% which is still lower than the efficiency achieved by Snowflake.

The accelerator by Qiu, et. al. \cite{qiu_embedded_fpga} is an FPGA-based implementation and is able to achieve 80\% efficiency on VGG16. The paper does not provide peak performance and this metric was computed using the same formula as Eyeriss. Measured performance is available and, like Eyeriss, is also split into performance with DRAM access latency and performance without DRAM access latency. Again, we consider the higher number that does not include DRAM access latency. It should be noted that Snowflake's performance and efficiency do take DRAM access latency into account. We consider this a fair comparison because Snowflake is able to completely hide DRAM latency behind the layer's processing by efficient use of traces and double buffering. Thus, our performance and efficiency with and without DRAM latency are the same.

The hardware convolution engine (HWCE) presented in \cite{meloni} is implemented on the same Zynq device as Snow-flake. The paper provides values for both peak and measured performance. Based on these, we computed efficiency for the HWCE as 88\%.

Compared to these, Snowflake is able to achieve 94\% for AlexNet, 91\% for GoogLeNet and 95\% for ResNet-50. Our lowest efficiency is still 3 percentage points above the next best result (HWCE) and is measured on a significantly more complex CNN model (AlexNet vs. GoogLeNet). Using a model to model comparison, Snowflake's efficiency is 6 percentage points higher than the HWCE. 

It should also be noted that none of these designs run GoogLeNet or any of the ResNet models as a benchmark. In fact, to the best of our knowledge, no implemented CNN accelerator has used GoogLeNet or ResNet-50 as a benchmark, despite both being more recent models than AlexNet or VGG. The Neurostream accelerator \cite{azarkhish_neurostream} provides numbers for newer models including GoogLeNet and ResNet but is a simulated design. They also require a significantly higher bandwidth of 32\,GB/s.. Finally, Neurostream's efficiency is obtainable when processing images in batch mode which is unsuitable for workloads that require low latency such as autonomous driving. Performance and efficiency when using a batch of 1 are not listed in the paper.

Snowflake is also able to achieve a higher clock frequency than the other designs. To the best of our knowledge, no FPGA-based implementation has achieved a clock frequency of 250\,MHz.

\section{Future Work and Conclusion}

This paper presented an efficiency and model agnostic CNN accelerator architecture called Snowflake. We implemented Snowflake on a Xilinx Zynq XC7Z045 device using 256 MAC units at 250\,MHz. Snowflake was able to achieve a throughput of 100 frames per second and 120\,G-ops/s on AlexNet and 36.4 frames per second and 116.4\,G-ops/s on GoogLeNet.

The Zynq XC7Z045 device has 900 MAC units. Scaling Snowflake up by using three compute clusters, we will be able to utilize 768 MAC units. Assuming an accelerator frequency of 250\,MHz, Snowflake will be able to achieve a peak performance of 384\,G-ops/s. Snowflake can be scaled further on larger FPGAs by increasing the number of clusters. If batch processing is used, computational efficiency will remain constant as the current system. Such a system can be used for server-based workloads where latency is not as important as throughput.

\bibliographystyle{ieeetr}
\bibliography{ref}

\begin{thebibliography}{10}

\bibitem{alexnet_owt}
A.~Krizhevsky, ``One weird trick for parallelizing convolutional neural
  networks,'' {\em CoRR}, vol.~abs/1404.5997, 2014.

\bibitem{googlenet}
C.~Szegedy, W.~Liu, Y.~Jia, P.~Sermanet, S.~Reed, D.~Anguelov, D.~Erhan,
  V.~Vanhoucke, and A.~Rabinovich, ``Going deeper with convolutions,'' in {\em
  The IEEE Conference on Computer Vision and Pattern Recognition (CVPR)}, June
  2015.

\bibitem{inception_v3}
C.~Szegedy, V.~Vanhoucke, S.~Ioffe, J.~Shlens, and Z.~Wojna, ``Rethinking the
  inception architecture for computer vision,'' {\em CoRR},
  vol.~abs/1512.00567, 2015.

\bibitem{resnet}
K.~He, X.~Zhang, S.~Ren, and J.~Sun, ``Deep residual learning for image
  recognition,'' {\em CoRR}, vol.~abs/1512.03385, 2015.

\bibitem{fcn_segmentation}
J.~Long, E.~Shelhamer, and T.~Darrell, ``Fully convolutional networks for
  semantic segmentation,'' in {\em The IEEE Conference on Computer Vision and
  Pattern Recognition (CVPR)}, June 2015.

\bibitem{resnet_segmentation}
Z.~Wu, C.~Shen, and A.~van~den Hengel, ``Wider or deeper: Revisiting the resnet
  model for visual recognition,'' {\em CoRR}, vol.~abs/1611.10080, 2016.

\bibitem{conneau_nlp}
A.~Conneau, H.~Schwenk, L.~Barrault, and Y.~Lecun, ``Very deep convolutional
  networks for text classification,'' {\em arXiv preprint arXiv:1606.01781},
  2016.

\bibitem{lecun_lenet}
Y.~LeCun, L.~Bottou, Y.~Bengio, and P.~Haffner, ``Gradient-based learning
  applied to document recognition,'' in {\em Intelligent Signal Processing},
  pp.~306--351, IEEE Press, 2001.

\bibitem{imagenet}
O.~Russakovsky, J.~Deng, H.~Su, J.~Krause, S.~Satheesh, S.~Ma, Z.~Huang,
  A.~Karpathy, A.~Khosla, M.~Bernstein, A.~C. Berg, and L.~Fei-Fei, ``{ImageNet
  Large Scale Visual Recognition Challenge},'' {\em International Journal of
  Computer Vision (IJCV)}, vol.~115, no.~3, pp.~211--252, 2015.

\bibitem{better_than_humans}
K.~He, X.~Zhang, S.~Ren, and J.~Sun, ``Delving deep into rectifiers: Surpassing
  human-level performance on imagenet classification,'' {\em CoRR},
  vol.~abs/1502.01852, 2015.

\bibitem{inception_v4}
C.~Szegedy, S.~Ioffe, and V.~Vanhoucke, ``Inception-v4, inception-resnet and
  the impact of residual connections on learning,'' {\em CoRR},
  vol.~abs/1602.07261, 2016.

\bibitem{DeepCompression}
S.~Han, H.~Mao, and W.~J. Dally, ``Deep compression: Compressing deep neural
  networks with pruning, trained quantization and huffman coding,'' {\em
  International Conference on Learning Representations (ICLR)}, 2016.

\bibitem{SqueezeNet}
F.~N. Iandola, M.~W. Moskewicz, K.~Ashraf, S.~Han, W.~J. Dally, and K.~Keutzer,
  ``Squeezenet: Alexnet-level accuracy with 50x fewer parameters and
  {\textless}1mb model size,'' {\em CoRR}, vol.~abs/1602.07360, 2016.

\bibitem{han2016eie}
S.~Han, X.~Liu, H.~Mao, J.~Pu, A.~Pedram, M.~A. Horowitz, and W.~J. Dally,
  ``Eie: Efficient inference engine on compressed deep neural network,'' {\em
  International Conference on Computer Architecture (ISCA)}, 2016.

\bibitem{han2015learning}
S.~Han, J.~Pool, J.~Tran, and W.~Dally, ``Learning both weights and connections
  for efficient neural network,'' in {\em Advances in Neural Information
  Processing Systems (NIPS)}, pp.~1135--1143, 2015.

\bibitem{chen_diannao}
T.~Chen, Z.~Du, N.~Sun, J.~Wang, C.~Wu, Y.~Chen, and O.~Temam, ``Diannao: A
  small-footprint high-throughput accelerator for ubiquitous
  machine-learning,'' in {\em Proceedings of the 19th International Conference
  on Architectural Support for Programming Languages and Operating Systems},
  ASPLOS '14, (New York, NY, USA), pp.~269--284, ACM, 2014.

\bibitem{shidiannao}
Z.~Du, R.~Fasthuber, T.~Chen, P.~Ienne, L.~Li, T.~Luo, X.~Feng, Y.~Chen, and
  O.~Temam, ``Shidiannao: Shifting vision processing closer to the sensor,'' in
  {\em Proceedings of the 42Nd Annual International Symposium on Computer
  Architecture}, ISCA '15, (New York, NY, USA), pp.~92--104, ACM, 2015.

\bibitem{zhang_caffeine}
C.~Zhang, Z.~Fang, P.~Zhou, P.~Pan, and J.~Cong, ``Caffeine: Towards uniformed
  representation and acceleration for deep convolutional neural networks,'' in
  {\em Proceedings of the 35th International Conference on Computer-Aided
  Design}, ICCAD '16, (New York, NY, USA), pp.~12:1--12:8, ACM, 2016.

\bibitem{qiu_embedded_fpga}
J.~Qiu, J.~Wang, S.~Yao, K.~Guo, B.~Li, E.~Zhou, J.~Yu, T.~Tang, N.~Xu,
  S.~Song, Y.~Wang, and H.~Yang, ``Going deeper with embedded fpga platform for
  convolutional neural network,'' in {\em Proceedings of the 2016 ACM/SIGDA
  International Symposium on Field-Programmable Gate Arrays}, FPGA '16, (New
  York, NY, USA), pp.~26--35, ACM, 2016.

\bibitem{alexnet}
A.~Krizhevsky, I.~Sutskever, and G.~E. Hinton, ``Imagenet classification with
  deep convolutional neural networks,'' in {\em Advances in Neural Information
  Processing Systems 25} (F.~Pereira, C.~J.~C. Burges, L.~Bottou, and K.~Q.
  Weinberger, eds.), pp.~1097--1105, Curran Associates, Inc., 2012.

\bibitem{vgg}
K.~Simonyan and A.~Zisserman, ``Very deep convolutional networks for
  large-scale image recognition,'' {\em CoRR}, vol.~abs/1409.1556, 2014.

\bibitem{nnx_by_moi}
V.~Gokhale, J.~Jin, A.~Dundar, B.~Martini, and E.~Culurciello, ``A 240 g-ops/s
  mobile coprocessor for deep neural networks,'' in {\em Proceedings of the
  IEEE Conference on Computer Vision and Pattern Recognition Workshops},
  pp.~682--687, 2014.

\bibitem{farabet_neuflow_somehow_still_novel}
C.~Farabet, B.~Martini, P.~Akselrod, B.~Corda, S.~Talay, Y.~LeCun, and
  E.~Culurciello, ``Bio-inspired vision processor for ultra-fast object
  categorization,'' in {\em Proc. High Performance Embedded Computing
  Workshop}, 2010.

\bibitem{reduced_precision_gupta}
S.~Gupta, A.~Agrawal, K.~Gopalakrishnan, and P.~Narayanan, ``Deep learning with
  limited numerical precision.,'' in {\em ICML}, pp.~1737--1746, 2015.

\bibitem{chen_eyeriss_isca}
Y.-H. Chen, J.~Emer, and V.~Sze, ``Eyeriss: A spatial architecture for
  energy-efficient dataflow for convolutional neural networks,'' {\em SIGARCH
  Comput. Archit. News}, vol.~44, pp.~367--379, June 2016.

\bibitem{chen_eyeriss_jssc}
Y.~H. Chen, T.~Krishna, J.~S. Emer, and V.~Sze, ``Eyeriss: An energy-efficient
  reconfigurable accelerator for deep convolutional neural networks,'' {\em
  IEEE Journal of Solid-State Circuits}, vol.~52, pp.~127--138, Jan 2017.

\bibitem{zhang_CNN_optimized_FPGA_arch}
C.~Zhang, P.~Li, G.~Sun, Y.~Guan, B.~Xiao, and J.~Cong, ``Optimizing fpga-based
  accelerator design for deep convolutional neural networks,'' in {\em
  Proceedings of the 2015 ACM/SIGDA International Symposium on
  Field-Programmable Gate Arrays}, FPGA '15, (New York, NY, USA), pp.~161--170,
  ACM, 2015.

\bibitem{meloni}
P.~Meloni, G.~Deriu, F.~Conti, I.~Loi, L.~Raffo, and L.~Benini, ``Curbing the
  roofline: A scalable and flexible architecture for cnns on fpga,'' in {\em
  Proceedings of the ACM International Conference on Computing Frontiers}, CF
  '16, (New York, NY, USA), pp.~376--383, ACM, 2016.

\bibitem{azarkhish_neurostream}
E.~Azarkhish, D.~Rossi, I.~Loi, and L.~Benini, ``Neurostream: Scalable and
  energy efficient deep learning with smart memory cubes,'' {\em CoRR},
  vol.~abs/1701.06420, 2017.

\end{thebibliography}

\end{document}